\documentclass{edm_template}
\usepackage{float}

\usepackage{tabularx, colortbl}
\usepackage{listings}
\usepackage{booktabs}
\usepackage{rotating}
\usepackage{multirow}
\usepackage{adjustbox}
\usepackage{graphicx}
\usepackage[stable]{footmisc}
\usepackage{hyperref}
\usepackage{makecell}
\usepackage{subcaption}
\usepackage{amsmath}
\usepackage{amssymb}

\usepackage{balance}
\usepackage[moderate]{savetrees}
\begin{document}

\title{Your Actions or Your Associates?\\ Predicting Certification and Dropout in MOOCs with Behavioral and Social Features
}
\numberofauthors{1} 
\author{
\begin{tabular}[t]{c@{\extracolsep{2em}}c} 
Niki Gitinabard, Farzaneh Khoshnevisan, \& Collin F. Lynch & Elle Yuan Wang\\
\affaddr{North Carolina State University} & \affaddr{Columbia University}\\
\affaddr{Raleigh, NC, USA} & \affaddr{New York City, NY, USA}\\
\email{\{ngitina, fkhoshn, cflynch\}@ncsu.edu} & \email{elle.wang@columbia.edu}\\
\end{tabular}
}

\maketitle
\begin{abstract}
    The high level of attrition and low rate of certification in Massive Open Online Courses (MOOCs) has prompted a great deal of research. Prior researchers have focused on predicting dropout based upon behavioral features such as student confusion, click-stream patterns, and social interactions. However, few studies have focused on combining student logs with forum data.
    In this work, we use data from two different offerings of the same MOOC.  We conduct a survival analysis to identify likely dropouts.  We then examine two classes of features, social and behavioral, and apply a combination of modeling and feature-selection methods to identify the most relevant features to predict both dropout and certification.  We examine the utility of three different model types and we consider the impact of different definitions of dropout on the predictors.  Finally we assess the reliability of the models over time by evaluating whether or not models from week 1 can predict dropout in week 2, and so on. The outcomes of this study will help instructors identify students likely to fail or dropout as soon as the first two weeks and provide them with more support.

    
\end{abstract}

\addtolength{\parskip}{-1in}

\section{Introduction}
Massive Open Online Courses (MOOCs) can provide broad and potentially scalable platforms for learning. Truly open MOOCs allow students around the world to enroll in any course that piques their interest or meets professional needs. Most of the available MOOCs are free, and many stay open perpetually even after their official offerings are complete thus allowing students to use them as a regular reference point or as a social platform. 

One major concern with MOOCs is that they have extremely high rates of dropout.  More than 85\% of students who register for a MOOC quit without completing it \cite{jordan14}.  Prior research has indicated that student dropout in MOOCs, and student performance more generally, is highly correlated with features of the students' online activities such as viewing lectures or attempting mastery quizzes \cite{pursel16, brooks15,joksimovic15, brown15, kloft14, sinha14, rose14, yang16, yang13, halawa14, fei15, andres16, chen17, eckles12, andres16, jiang14a,jiang14b}. These activities can be classified as student-system interactions (e.g. video viewing) \cite{pursel16, brooks15, kloft14} and student-student interactions (e.g. posting to a forum) \cite{joksimovic15, brown15, rose14,eckles12, andres16, jiang14a,jiang14b}. 

Social network analyses of interactions among students has shown that students' social interactions and social presence metrics can be used to predict their performance \cite{joksimovic15, brown15, rose14, zhu16}. However, in most of these studies, the authors did not focus on how the students form their social networks over time. Nor did they examine whether or not the different types of user forums produced substantively different networks. Similarly, prior studies of dropout prediction from activity logs have shown that students' study habits can be used to predict attrition \cite{fei15, yang16, yang16, andres16, sinha14}. However activity logs and social metrics cover very different aspects of student behavior.  Therefore it is possible that by combining the two, we may be able to improve our insights into students' behaviors and thus, improve our ability to predict both performance and dropout. 
Few researchers have combined behavioral and social metrics to improve prediction performance \cite{fei15, taylor14}. Thus it is beneficial to make this comparison on new datasets to check the generality of the outcomes. 

Prior researchers have also shown that students' actions during the first few weeks of a course can be used to predict their subsequent performance \cite{brooks15, fei15, kloft14, taylor14, jiang14b}. It has also been shown that models trained on one class can sometimes be applied to other classes \cite{brooks15, vihavainen13, boyer15}, but these findings have only been tested on a few MOOCs and are not yet reliable. Therefore it is an open question whether metrics of the type that we consider will be transferable.

In this study, we used two different offerings of a MOOC on Big Data in Education, offered by Dr. Ryan Baker on the Coursera Platform in 2013 and EdX in 2015. We generated social networks based upon two approaches taken by the prior studies for the same dataset based on different sets of assumptions, compared them and show how changing assumptions can affect the findings and also, how forum structure can help us make assumptions with more caution \cite{zhu16, brown15}. We also perform a survival analysis to find the groups of students that are more likely to dropout and compare the findings among both classes. Then we use feature selection to find out which features can provide us with more information gain. Later, we train predictive models using the top features in our feature selection and predict dropout and certification. Finally, we use the prediction models trained on each week of the first offering of the earlier course to predict dropout and certificate earning early in the second offering. 

Overall, we aim to investigate the following research questions:
\vspace{-0.6cm}
\begin{enumerate}
    \item What features are most predictive of student drop-out?
\item How will the choice of target label, social graph generation, and features affect prediction results?
\item How early can we predict student dropout in MOOCs?
\item Can we make predictions across course offerings by using models trained on one year to predict others?
\end{enumerate}

As part of this work we will show how important the assumptions we make are on the performance and findings of the study. The generated models can also help MOOC instructors identify students who are likely to dropout early in the semester using models from prior classes and provide the students with more support and motivation to complete the course.
\begin{table*}[h]
\begin{center}
\begin{adjustbox}{width=0.97\textwidth}
    \begin{tabular}{lccccccccccc}
    \hline
      Data & \makecell{Enrolled\\ Students} & \makecell{Forum Active\\ Students} & \makecell{Students Who\\ Had Some Submissions} & \makecell{Number of\\ Forum Posts} & \makecell{Students Who \\Had Some Activity} & \makecell{Non-zero\\ Grades} & \makecell{Earned\\ Certificates} & \makecell{Thread \\Count} & \makecell{Thread Avg \\Length} & \makecell{Thread Max \\Length} & \makecell{Thread Min \\Length} \\
      \hline
        BDE 2013 & 55,013 & 750 & 1,599 & 4,261 & 17,295 & 1,381 & 638 & 281 & 5.31 & 89 & 1\\
        BDE 2015 & 10,190 & 519 & 1,437 & 2,063 & 5,077 & 320 & 117 & 624 & 2.24 & 36 & 1\\
        \hline\\
    \end{tabular}
    \end{adjustbox}
    \caption{BDE MOOC 2013 and 2015 Characteristics, Including Only the Students Who Started and Finished the Course On-schedule}
    \label{tab:data_stats}
    \end{center}
\end{table*}
\section{Background}
Prior research on 39 MOOCs showed that on average only 6.5\% of the users who enroll in a MOOC finish with a passing grade and earn a certificate \cite{jordan14}. As Yang et al. noted, this high attrition may be caused by several factors such as students losing interest  over time, or by mounting confusion and frustration.  Or it may simply be the case that they never intended to complete the course in the first place \cite{yang16}. We acknowledge that some users enroll in MOOCs only to access specific parts of the material and with no intention of obtaining a certificate and that intention to finish the course is correlated with course completion \cite{pursel16, gutl14}. Pursel et al., for example, showed that students' plans to watch videos and earn a certificate is a significant predictor of their course completion \cite{pursel16}. Gutl et al. surveyed students who did not complete a course and found that only 22\% of them had intended to do so in the first place \cite{gutl14}. 

In addition to intentions and motivation, researchers have also observed that other attributes are useful for predicting students' course completion. These include: the number of videos that a student watches in a week; the number of quiz or assignments they attempt; the number of forum posts made per week along with the post length; the time spent on assignments; whether they spend more time on forums or on the assignments; whether or not they start early; and demographic data such as their age, fluency with English, and their education level \cite{pursel16, fei15, rose14, yang13,andres16, chen17, sinha14}. 
Some researchers have also utilized social network metrics such as degree, centrality, hub, and authority scores \cite{gitinabard17w, jiang14a, yang13, joksimovic15, brown15, rose14, zhu16}.

Joksimovic et al. showed that students' social presence metrics can be used to predict their final grades \cite{joksimovic15}. Some examples of these parameters include: continuing a thread, complimenting other users, and expressing appreciation.  Eckles et al. went further than general graph attributes and observed that whether or not a students' best friend stays in the course is strongly correlated with whether or not they do so \cite{eckles12}.  Unlike other researchers, Eckles et al. did not use a social network to define this relationship but surveyed the students directly. Brown et al. analyzed the same 2013 dataset that we use here.  They showed that students form communities based on their interactions on the discussion forum and membership of these groups are correlated with the students' grades \cite{brown15w}. In the prior literature, different methods have been used to generate social networks, but few comparative studies have been done to highlight their effects. Brown et al. \cite{brown15w} and Zhu et al. \cite{zhu16} exemplify some of the alternatives. Brown et al. formed a weighted undirected social network by connecting each author that posts to a discussion thread with all of the authors that had previously contributed to it, on the assumption that each author reads the current thread before adding to the conversation and that the reply is intended for all authors \cite{brown15}. Thus, the graph assumes an implicit social connection by virtue of the group conversation. Zhu et al., by contrast, added a connection from each author who contributes to a thread to the author of the first post alone on the assumption that the thread consists of a series of flat replies to the original post and that users will only read the first post before replying \cite{zhu16}. Whether or not these assumptions are valid depends upon the structure of the forums and the habits of the students themselves. Indeed they depend upon the ``culture'' of the class. It is therefore important to study the impact of these assumptions on the outcome of a study.

Prior research has shown that these predictive models can not only be used to predict students' performance based upon the data from the entire semester in the same class, that they can also be used to make early predictions, based upon partial class data, or to make predictions across classes. Previous studies used a model trained on one offering of a MOOC to make predictions for another \cite{boyer15, brooks15}. An early notifier to identify student performance in the course using only the first few weeks of data in MOOCs has also been investigated before \cite{brooks15, jiang14b}.


As prior research shows, both behavioral and social features are predictive of dropout. These features cover different aspects of student activities, we therefore decided to use a selection of both types of features to train our predictive models. Fei et al. used a combination of these features to generate predictive models, but they did not evaluate this hybrid approach against pure activity or social models \cite{fei15}. Taylor et al., however, has shown that in their MOOC, the addition of forum activity did not add much value to a previous log-based predictive model \cite{taylor14}. It is therefore important to study whether or not combining these feature types can make a difference in different courses because it might depend on the course structure and its use of the discussion forum.

\section{Dataset}
We analyzed data from two different offerings of the ``Big Data in Education'' MOOC (BDE MOOC), from 2013 and 2015. Table \ref{tab:data_stats} presents some basic characteristics of these two datasets. The presentation and storage formats were slightly different as in 2013 it was offered on the Coursera platform while in 2015 it was deployed on EdX. We will therefore describe them separately in the following sub-sections.

``Big Data in Education'' course was offered by the Teacher's College at Columbia University on the Coursera platform in 2013. A total of 55,013 students registered for the course, but only 17,295 had any activity recorded in the logs. Only 750 students made one or more posts or replies on the discussion forum. Our dataset does not include view records so we cannot estimate how many students visited the forum but made no contribution. Roughly 1,599 students submitted assignments or quizzes. In this study, we considered 23,080 students who had at least a recorded activity in the forum, assignment submission, or lecture view. Both of the courses were open for students after the official offering was over.  Therefore the datasets included students who worked on their own well after the instructor and the rest of the class had left. For this analysis we decided to focus solely on those students who started and finished the courses on schedule so that their activities would fit properly into the official weeks and the course calendar. This left  a total of 17,295 students remaining in our dataset. We extracted the grades for these remaining students. Among all, only 1,381 had non-zero final grades.

This class was offered again on the EdX platform in 2015. As before, the provided data consisted of activity logs, final certificates, and forum posts. In addition to the threaded discussion forum, edX also offers a chat platform among participants of the course called Bazaar where a lot of the discussions among students take place. Unfortunately, the data from that platform was not available for this study. A total of 10,190 students were initially enrolled in this class. Only 519 students posted or replied on the forum, 1,437 submitted at least one of the problems, and 320 students had a non-zero final grade. As with the 2013 dataset, we removed the students who had submissions before or after the course dates leaving 5,077 students.\\

\section{Methods}
We began our analysis by generating a social network, and extracting structural and behavioral features from it and the logs.  We ran feature selection to determine whether or not a combination of these features can improve the performance of the overall model, when compared to using each of the groups separately. In the final step of this process we rah a machine learning analysis to predict dropout and certification. We extracted each of the features on a week-by-week basis.  Thus we produced a set of per-week datasets each of which includes all data before the end of the associated week. This will help us to analyze how early we can predict dropout and certificate earning based on their activities so far. We will discuss each of these steps in the following subsections.\\

\subsection{Graph Generation}\label{social_features}
\begin{figure}[h]
  \centering
  \begin{subfigure}[h]{0.35\textwidth}
        \centering
        \includegraphics[height=0.9in]{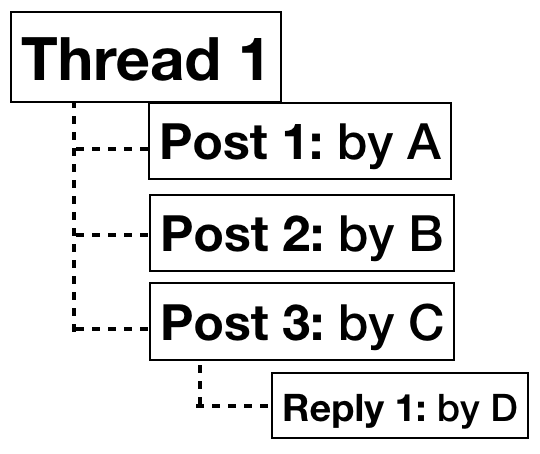}
        \caption*{(a)}
  \end{subfigure}
  \begin{subfigure}[h]{0.2\textwidth}
        \centering
        \includegraphics[width=0.6\linewidth]{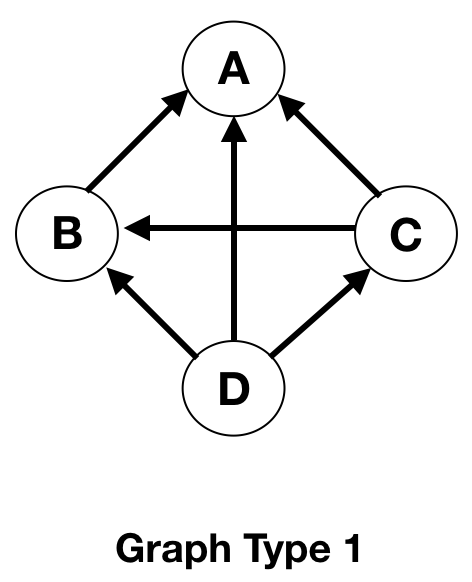}
        \caption*{(b)}
    \end{subfigure}
    ~
    \begin{subfigure}[h]{0.2\textwidth}
        \centering
        \includegraphics[width=0.6\linewidth]{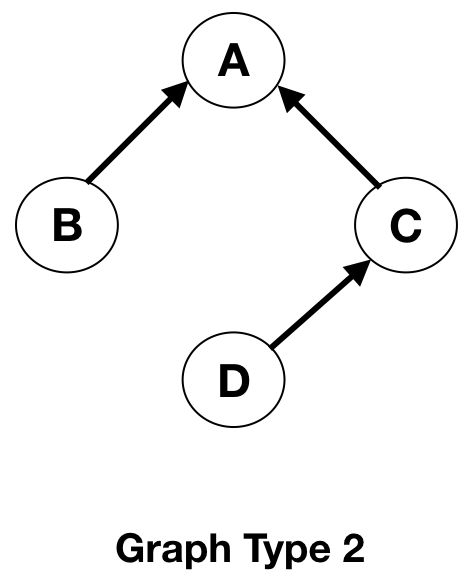}
        \caption*{(c)}
    \end{subfigure}
    
  \caption{Graph construction of Type 1 and 2 for post/reply structure example}
  \label{fig:graph_example}
\end{figure}

In both classes the forums consist of a series of threaded discussions. Class participants may initiate a thread by making a root post and may reply to existing threads by adding comments at the end or by replying to a specific post. As mentioned above, two approaches have been used to generate social graphs from discussion forums. Brown et al. connected authors of all the posts and replies in a thread to authors of all the preceding contributions in the thread \cite{brown15}. This method assumes that everyone who posts on a thread or replies to a post has read all of the preceding posts on the same thread first and is responding to all of them. Another approach used by Zhu et al. suggests connecting all the authors in a thread to the author who originated it \cite{zhu16}. This approach is more reasonable for flat forums where  each thread is a separate question and all of the replies are directed towards the first post. In this study, we generated the social graphs based upon both approaches.  We designate Brown's approach ``Type 1'' and Zhu's approach ``Type 2''. Figure \ref{fig:graph_example} shows an example of a thread structure and the two corresponding graphs to highlight differences between these methods. We then compared these graphs in terms of their ability to predict both dropout and whether or not students would earn a certificate, only among those who lie on the graph. The structures of the forums differ between Edx and Coursera, so we expect that this difference will be reflected in the relative performance of the classifiers on these graphs. On the Coursera platform, used for BDE 2013, clicking on the first post in a thread will show all of the remaining posts as well as replies to them.  Thus it makes sense to construct a Type 1 graph and to connect every author to the authors of the preceding posts. However, the structure of the EdX forum is slightly different. Once a thread is selected, you see the beginning of all the posts but not the full text. By selecting each post you can view the full content and the replies. Therefore, when reaching a specific post, the users do not necessarily need to view preceding comments. In this case, it seems more reasonable to construct a Type 2 graph by connecting replies to the original post alone. 


The length and the number of threads for each class is shown in Table \ref{tab:data_stats}. In 2013 there were fewer threads than in 2015 but the threads themselves were generally longer.  This may be a consequence of the difference in the platforms, the nature of the discussion forums, the addition of the chat platform, or how the users learned to interact with the tools.




Since we are focused on student to student interactions we chose to remove the instructor from the graphs. We also removed all of the isolated nodes (students who did not make posts or receive replies) before calculating the social metrics as all metrics for an isolated node would be zero. 

\subsection{Generated Features}\label{features}
For each student in the graphs we calculated the following features: \textit{Betweenness Centrality} showing to what extent a vertex lies on the paths between other users \cite{freeman}, which indicates the importance of the student in connecting other students together; \textit{Hub score} showing the extent that a node points to many good authorities \cite{kleinberg}, students with higher hub scores, respond to active students' posts more frequently; \textit{Authority score} showing the extent that a node is pointed by many good hubs \cite{kleinberg}, students with higher authority scores, receive comments from hub students more frequently; \textit{In-degree} showing the number of connections a student has received by getting replies from others; \textit{Out-degree} showing the number of connections the student has made by posting replies to others; and \textit{Dropped\_out\_neighbors} showing the proportion of a user's neighbors that have already dropped out in each week. This metric was inspired by Eckles et al. \cite{eckles12}, and was defined as a way to estimate whether or not the students' attrition can be affected by their closest neighbors. This feature can show how much a user has been exposed to unmotivated users.

In addition to the social features described above, we defined other general features based upon students' log data and forum activity. Some of these, which we call \emph{forum features}, are based on activities on the forum including the \textit{total\_posts}, \textit{total\_comments}, as well as the total number of \textit{votes} (total upvotes $-$ total downvotes) that the student received on their posts.
The third group of features, called \textit{behavioral features}, is extracted from the activity logs.
We extracted the total \textit{video\_view}s and \textit{video\_download}s for the 2013 students class. The 2015 offering did not provide download information. In 2015 students were offered `chapters' to view. We therefore extracted the total number of \textit{video\_views} and \textit{chapter\_views} for this class. The \textit{total\_attempts} is also included in both cases.  This represents the total number of assignment submissions for each student. 

The last group of extracted features, which is our target for prediction, includes \textit{semester\_dropout}, \textit{week\_dropout}, \textit{inactive\_next\_ week}, and \textit{certificate}. Defining dropout based on observations of online activities is not trivial because the students do not explicitly declare their leaving. Prior studies have proposed different measures reflecting dropout \cite{fei15, yang16}. We define our measures similar to Fei et al. as described below and generate our predictive models based on all of them \cite{fei15}. Mostly our focus in this paper will be on semester dropout and certificate earning because they provide a static label for students over all the weeks of the semester.\\
\textbf{Semester dropout}: Will this student stop engaging at some point? This feature is represented by a boolean flag which indicates that the student dropped out of the course \emph{before the end}.  Thus if a student quits performing actions in the course in any week but the last then this will be set to 1 for all weeks.  We do not consider students with no activity in the last week as dropout since they may have finished earlier in the week.\\
\textbf{Week dropout}: Will this student stop working from next week? This is a boolean flag that is used to designate when a student drops out. It will be set to 1 for a week if the student does not perform any activities in the subsequent weeks. The activities we consider include: posting or commenting on the forum, submitting assignment, and watching or downloading lecture videos (or chapter view in BDE 2015 data).\\
\textbf{Inactive next week:} shows whether the student will be inactive in the following week.\\
\textbf{Certificate:} shows whether the student has earned a certificate.

\subsection{Survival Analysis}
Survival analysis is the analysis of data involving the time remaining to the occurrence of some event of interest. This method was originally introduced in medical research and is used to predict how long patients would survive, or go without some change, based upon their data \cite{miller11}. It has since been adapted to a number of other fields where estimating the time until the occurrence of an event or a boolean flag is of interest \cite{yang16}. One objective of survival analysis is to examine whether the survival times are related to other features. For this purpose, regression models can be used to assess the effect of covariates on an outcome. In this study we used a multivariate version of Cox proportional hazards model to fit the hazard ratio at time $t$ as follows:

\vspace{-1.5em}
\begin{equation}
    h(t;x_1,...,x_n) = h_0(t)e^{\beta_1x_1+...+\beta_nx_n}
\end{equation}
\vspace{-1.5em}

In this formula, $h_0(t)$ is the baseline hazard that is the hazard ratio of an individual, at time $t$ where all the covariates are zero. The effect of variable $x_k$ while all other variables are fixed is interpreted as: for each unit increase in $x_k$ with all other variables held fixed, the hazard is multiplied by $e^{\beta_k}$ \cite{kartsonaki2016survival}. 

Here, we used the week of dropout as the target time. For the students who have not dropped out until week 6, we consider them as not dropping out, via right censoring. \emph{Right censoring} occurs when an individual has not had the event of interest until the end of study. 
Further, we normalized all the variables to have a mean of zero and a standard deviation of one. Therefore, the resulting hazard ratio of 1.2 for total number of attempts for example, shows that students with one standard deviation higher attempts than average are 20\% more likely to dropout. Likewise, a hazard ratio of 80\% would show the users are 20\% less likely to dropout.
For this study, we generated two models, the first included social features alone while the second one included all of the predefined features.

\subsection{Dropout Prediction}
As mentioned earlier, we generated multiple datasets for each class, each of which corresponds to the activities up to the end of a given week, starting from the first week of the course. We then built classifiers for each dataset to evaluate their performance. Our primary goal was to determine how early the data would be sufficient for the model to train and to predict the outcome with high accuracy. In both classes, the percentage of students who stayed engaged in the course until the end was less than 20\% and the ones who earned a certificate were around 2\%, so a model trained on that data would generate biased results. To make the dataset balanced, we randomly removed some of the majority class instances until the number of users in both the classes were equal. For the prediction, we began by employing decision tree-based feature selection based on Gini impurity index to select the most important features for the prediction of each target class \cite{dash1997feature}. We then applied Support Vector Machine (SVMs) and Logistic Regression (LR) for the classification task. Since Logistic Regression outperformed SVM in all cases, we only report the AUC performance of the logistic regression when comparing different models. In order to tune the parameters and validate the training procedure, we applied 10-fold nested cross-validation to estimate each outcome. 

\section{Results and Discussion}

\subsection{Graph Construction}
The construction method used when making a social graph can affect the predictive performance. In this section, we assess the predictive power of the graph features, for both dropout and certification, that we extracted from the two different social graph types.
Table \ref{tab:graph_cpmparison} presents the predictive performance of an LR classifier trained on the aforementioned features extracted from the two graph types for BDE 2013 class. As we hypothesized, for both semester dropout and certificate prediction, \textit{Type 1} graph features perform slightly better than the \textit{Type 2}. 

\begin{table}[htb]
\begin{adjustbox}{width=0.47\textwidth}
    \centering
    \begin{tabular}{clcc|cc}
    \hline
        \multirow{2}{*}{Class} & \multirow{2}{*}{Target} & \multicolumn{2}{c|}{AUC} & \multicolumn{2}{c}{F-measure}\\
        &  & Graph 1 &Graph 2&Graph 1&Graph 2\\ \hline
        \multirow{2}{*}{BDE 2013}& Semester Dropout & 0.72&	0.707&	0.709&	0.689\\
        & Certificate & 0.808&	0.796&	0.763&	0.749\\ \hline
        \multirow{2}{*}{BDE 2015}& Semester Dropout & 0.548&	0.577&	0.559&	0.666\\
        & Certificate & 0.545&	0.607&	0.491&	0.676\\
        \hline
    \end{tabular}
\end{adjustbox}
\caption{Graph Construction Effect on Dropout and Certification}
\label{tab:graph_cpmparison}
\end{table}

Table \ref{tab:graph_cpmparison} shows the same comparisons for the BDE 2015 class offered on the EdX platform. Consistent with our hypothesis, the predictive model based on graph 2 features performs considerably better than the graph 1-based model. Thus, we will use this construction of the graph for dropout and certificate prediction later. Overall, the graph features in BDE 2015 are less predictive of student outcomes than with the BDE 2013 data.  Similar to the difference in the length of the threads, this may also be due in part to the presence of the separate chat platform where part of the discussion among students takes place. Those interactions are not represented in our dataset.
As our results show, the methods and assumptions used for the generation of social graphs should be tailored to the class and forum structure and some methods may not generalize to all of the other classes or platforms.

\subsection{Survival Analysis}
We explored two different sets of features in our survival analysis to discover the impact of the features on dropout in the two course offerings. In BDE 2013, when including all the aforementioned features in Section \ref{features}, we observed that both the behavioral and social features have a high hazard ratio and significant p-values as shown in Table \ref{tab:survival_2013_g1}. Accordingly, the hazard ratio (HR) 0.71 for video download indicates that students who download one standard deviation (SD) more videos than average are 29\% less likely to dropout compared to the ones with an average number of video downloads. 
Betweenness with a hazard ratio of 1.74 illustrates that the students with one SD more betweenness than average are 74\% more likely to dropout. We examined some sample posts made by the students with high betweenness. It appears that many of the posts are social niceties such as expressions of gratitude or  appreciation for the instructor or fellow students rather than being substantive contributions to the discussion (e.g. ``Nice work'' or ``Your kind of persistence will always pay off eventually'').  

In our social model, we only considered the features that were extracted from the social graph in order to assess their effect on students' survival in the course. As our results suggest, the students whose out-degree or in-degree are one SD higher than the average are 22\% and 40\% less likely to dropout respectively. This means that the students who typically answer others' questions or post new questions are more likely to stay active in the course. When comparing this finding with betweenness from the previous model, we can conclude that the students with only high in-degree might be more confused, while the students with high out-degree probably understand the material better, or think that they do, and are more willing to share information.  Doing both at the same time however, may show that the student is  interested in socializing rather than information exchange, which may not help them to understand the material, complete the course, or gain a certificate because the socialization may take priority over learning.

\begin{table}[h]
\begin{adjustbox}{width=0.47\textwidth}
    \centering
    \begin{tabular}{l|cc|lc|lc}
    \hline
        \multirow{2}{*}{Features}&& & \multicolumn{2}{c|}{No grade} & \multicolumn{2}{c}{Social}\\
        &Mean&SD& HR & SE & HR & SE\\
        \hline 
            video\_download&14.46&68.25 & 0.71***	&0.03&\multicolumn{2}{c}{---}\\
            total\_attempts&0.66&3.27 & 0.57***&	0.04&\multicolumn{2}{c}{---}\\
            total\_posts&0.03&0.34&	0.63***&	0.13&\multicolumn{2}{c}{---}\\
            indegree&0.27&2.67&	0.75**&	0.09&	0.60***&	0.10\\
            outdegree&0.27&2.68 & &&	0.78*&	0.09\\
            betweenness&17.58&326.42&1.74***&	0.14&&\\
        \hline
    \end{tabular}
\end{adjustbox}
\caption{BDE MOOC 2013 - Survival Analysis for Different Models {\scriptsize(*: $p < 0.05$, **: $p < 0.01$, ***: $p < 0.001$, ---: not included)}}
\label{tab:survival_2013_g1}
\end{table}

The survival analysis results for the BDE 2015 course is shown in Table \ref{tab:survival_2015_g3}. The strongest features in this offering are largely behavioral features such as chapter views, total posts, and the total number of attempts. Chapter views had a hazard ratio of 0.53.  Thus students with 1.5 more views than 2.32, are 47\% more likely to continue in the course.
In this case, the social features are not significant unlike the 2013 class. Additionally, having more posts in the 2013 class seemed to help people complete more, while in this class it had a negative influence. Comparing the instructor and TA activity in both classes shows that the instructor and the most active TA made many more comments in 2013 than in 2015. In 2013, the instructor and the most active TA made a total of 432 comments, while in 2015 only 133 comments were made by the instructor, and we identified no TA with significant activity. If we assume that most of the posts were expressions of confusion, the more replies that they received, the more likely it is for their confusion to get resolved. Based on the observed reply behavior of the teaching staff in those classes, it seems likely that confused students had a better chance of finding an answer in the 2013 class than in 2015. It is also possible that part of the support was provided to students via the separate chat platform, but in either case it seems that posting on the forum was less helpful in 2015 than 2013. 
This may indicate that posts and replies did not resolve confusion. Additionally, the results of the survival analysis align with the results of the comparison among predictive models presented in Table \ref{tab:graph_cpmparison}.


\begin{table}[h]
\begin{adjustbox}{width=0.47\textwidth}
    \centering
    \begin{tabular}{l|cc|lc|lc}
    \hline
        \multirow{2}{*}{Features}&& & \multicolumn{2}{c|}{No grade} & \multicolumn{2}{c}{Social}\\
        & Mean&SD&HR & SE & HR & SE\\
        \hline 
            chapter\_view&2.32&1.57 & 0.53***&	0.03&\multicolumn{2}{c}{---}\\
            total\_posts&0.2&1.31& 1.43*&	0.14&\multicolumn{2}{c}{---}\\
            total\_attempts&1.36&3.27& 0.88**&	0.04&\multicolumn{2}{c}{---}\\
            outdegree&0.02&0.26& &&0.43***&	0.18\\
        \hline
    \end{tabular}
\end{adjustbox}
\caption{BDE MOOC 2015 - Survival Analysis for Different Models {\scriptsize(*: $p < 0.05$, **: $p < 0.01$, ***: $p < 0.001$, ---: not included)}}
\label{tab:survival_2015_g3}
\end{table}

\subsection{Feature Selection}
The five most important features for each prediction task and their importance scores for BDE MOOC 2013 is shown in Table \ref{tab:2013_feature_selection}. As the dropout feature selection results show, video download, video view, and total attempts are the most important features for prediction of the semester dropout, while total posts and indegree are significantly less important and the remainder of the features do not show up. Therefore, we used the top three features to train our semester dropout classifier. Furthermore, when predicting certification, total attempt and video view features had the highest importance score and we used them for training the model. Similarly, in 2015, video view, chapter view, and total attempts had the highest importance score for both dropout and certificate prediction.

\begin{table}[htb]
\begin{adjustbox}{width=0.49\textwidth}
    \centering
    \begin{tabular}{lcc|cc}
    \hline
        &\multicolumn{2}{c|}{Semester Dropout}&\multicolumn{2}{c}{Certificate} \\ \hline \hline
        Rank &Feature & Importance & Feature & Importance\\ \hline \hline
        1& video\_download& 0.604& total\_attempts& 0.692\\
        2& video\_view& 0.230&video\_view&0.178\\
        3& total\_attempts&0.111&votes&0.045\\
        4& total\_posts& 0.013&indegree&0.038\\
        5& indegree& 0.011& total\_posts&0.019\\
        
        \hline
    \end{tabular}
\end{adjustbox}
\caption{BDE MOOC 2013 - Feature selection using Decision Tree}
\label{tab:2013_feature_selection}
\end{table}

        

Our observations showed that none of the forum features representing participation were as informative as the behavioral features. This was due in part to the fact that there was a small proportion of students who had any forum activity. 
Additionally, we have access to a survey completed by students before starting the course. A total of 155 and 229 students from the 682 and 483 forum active ones participated in the survey respectively in the 2013 and 2015 classes. An analysis on the responses of the forum active students shows that more than 66\% of them indicated the reason for taking the class as it being relevant to their field of study, more than 77\% of them indicated that it is relevant to their career, more than 87\% of them believed that it will help them expand their knowledge of the field, and only less than 40\% mentioned that it will help their resume. So, it seems like not many of them were concerned about finishing the course, getting a certificate, and using it as a boost to their resume. More information on the structure of the survey is available in Wang et al. \cite{wang14}. 
Also, some more analysis on the student replies to the survey and their certificate earning is available in Andres et al. \cite{andres16}.

\subsection{Model Performance}
\begin{figure}[h]
  \centering
  \includegraphics[width=0.47\textwidth]{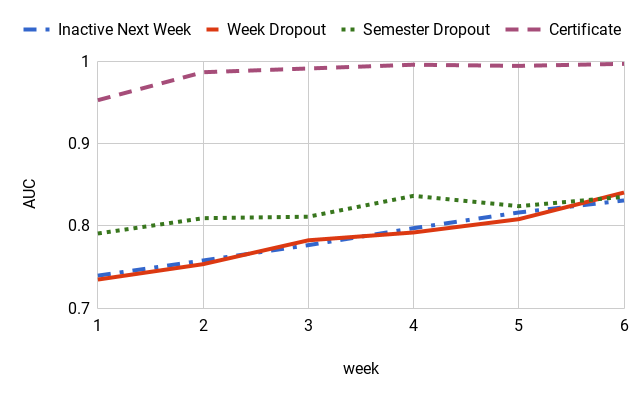}
  \caption{BDE 2013 - AUC Prediction Performance of Three Dropout Targets and Certificate}
  \label{fig:2013_model_comparison}
\end{figure}

To train our models, we only considered features with more than a 0.1 importance score in feature selection and applied logistic regression with 10-fold cross-validation to evaluate the model. Figure \ref{fig:2013_model_comparison} presents the AUC performance of each classifier over the first six weeks of the course. F-Measure performance also had a similar trend. As we observed, the certificate prediction model had an AUC above 90\% from the first week of the course.  While the model for semester dropout obtained an AUC of approximately 79\% in the first week and gradually increased thereafter. The Week dropout and inactive next week models behaved similarly.

The classification performance of the models for the BDE 2015 dataset was similar to 2013. As with the BDE 2013 dataset, the certificate prediction performance is above 85\% from the first week and improves gradually thereafter. The three dropout definitions have almost the same trend and behavior while obtaining an F-measure performance around 70\%.

Our results show that even though most of the more complicated metrics were removed during the feature selection method, the trained models for both of the classes were able to predict dropout with an F-Measure and AUC of  $\approx 70\%$ as early as the third week of the class. They were also able to predict certification with an F-Measure and AUC of $\approx 90\%$ from the very first week. It also appears that the students who earn a certificate have relatively distinct behavior from that of their peers starting at the beginning of the semester. 

It is interesting to note that some of the features showed significant outcomes in the survival analysis, but were not selected by the feature selection algorithm. This was surprising as one argument that has been advanced for survival analysis is that it is better at handling time-aware events. In order to evaluate this apparent conflict we trained a separate model based on the features that were significant for the survival analysis. The resulting model did not outperform the models that relied on the tree-based approach. In most cases the resulting models were comparable or the survival model underperformed. While this does not prove that survival-based selection is unusable it does merit further study.


\subsection{Cross-Class Dropout and Certification}
In order to evaluate the generalizability of the classification models, we took a cross-class approach. To do so, we used all of the data from each week of the first offering, based on the behavioral features common among both classes, which includes only video view and total attempts. Then we tested the model on all data from the corresponding week of BDE 2015. Table \ref{tab:cross_class} presents the F-measure and AUC performance of this model over six weeks of the course for the task of semester dropout and certificate prediction. As the results suggest, the predictive power of this model is relatively high when using only the two aforementioned features, especially for the certificate prediction task. This finding suggests that even though there are some differences among these classes and the features selected for the classification of each might be slightly different, the models are able to predict students' outcome as early as the first one or two weeks with reasonable accuracy. However, these results need to be validated on other courses with multiple offerings.

\begin{table}[H]
\begin{center}
\begin{adjustbox}{width=0.30\textwidth}

    \begin{tabular}{ccc|cc}
    \hline
        &\multicolumn{2}{c|}{Semester Dropout}&\multicolumn{2}{c}{Certificate} \\ \hline \hline
        Week &F-measure & AUC & F-measure & AUC \\ \hline \hline
        1&	0.606&	0.764&	0.727&	0.885\\
        2&	0.723&	0.883&	0.776&	0.935\\
        3&	0.706&	0.882&	0.741&	0.879\\
        4&	0.627&	0.879&	0.721&	0.871\\
        5&	0.490&	0.915&	0.750&	0.861\\
        6&	0.586&	0.915&	0.836&	0.951\\
        \hline
    \end{tabular}

\end{adjustbox}
    \end{center}
\caption{AUC and F-measure performance of Cross-class dropout and certificate prediction model}
\label{tab:cross_class}
\end{table}

\section{Conclusion}
Our primary goal in this study was to predict student performance and dropout based upon different social and behavioral features. One focus of our work here was on the testing assumptions that are usually made when generating social features and choosing the analysis method. These findings suggest that even for similar classes with the same instructor, a change in platform or instructor/TA behavior can change the impact or appearance of student engagement in the forums. As a result, the choice of model features and the feature generation methods matter a great deal. For example, we tried two different social graph generation methods that both were suggested in the literature and based on the forum structure the better choice for each class was different. Additionally, as our results suggest, behavioral features such as submissions and video watching are better predictors of student dropout and certification than social behavior. Adding social metrics to the trained behavioral models does not seem to improve their performance because very few users seem to place any value on those features. Additionally, we observed that a behavior-based predictive model trained on a former offering is applicable to a new offering, despite the differences in course structure. This suggests that we may be able to generate predictive models based upon early offerings of MOOCs and then use them on to enhance the later iterations. This will enable instructors to identify students who are likely to earn a certificate or to dropout in the first few weeks of the course and may be able to help or provide more support for the students in need.

One limitation of this work is that dropout is not pre-defined in the dataset, and we have no ground truth on the students' intentions when they quit.  We therefore need to make assumptions when defining dropout, which can change the findings of the study. Our definitions of dropout, would count students with any kind of activity still engaged, thus if a student kept watching videos but not submitting assignments, they would not be counted as having dropped out. However, different assumptions on the definition of dropout might change the findings. Also, our analysis of the posts made by the central users is limited. Deeper study of the content in posts and replies, or whether they are on topic, can make the findings stronger. 

One other limitation of our study was the imbalanced nature of our dataset. In both offerings, the majority of the students dropped out according to our definitions.  In order to address this problem we randomly removed most of the dropped out students to balance this label as half true and half false. In future studies, more approaches should be tried to balance the dataset, and to include more variety of data while removing the duplicates.

\section{Acknowledgments}
This work was supported by NSF grant \#1418269: ``Modeling Social Interaction \& Performance in STEM Learning'' Yoav Bergner, Ryan Baker, Danielle S. McNamara, \& Tiffany Barnes Co-PIs.

\clearpage
\balance
\bibliographystyle{abbrv}
\bibliography{references}

\begin{thebibliography}{10}

\bibitem{andres16}
J.~M.~L. Andres, R.~S. Baker, G.~Siemens, C.~A. Spann, D.~Ga{\v{s}}evi{\'c},
  and S.~Crossley.
\newblock Studying mooc completion at scale using the mooc replication
  framework.
\newblock 2016.

\bibitem{boyer15}
S.~Boyer and K.~Veeramachaneni.
\newblock Transfer learning for predictive models in massive open online
  courses.
\newblock In C.~Conati, N.~Heffernan, A.~Mitrovic, and M.~F. Verdejo, editors,
  {\em Artificial Intelligence in Education}, pages 54--63, Cham, 2015.
  Springer International Publishing.

\bibitem{brooks15}
C.~Brooks, C.~Thompson, and S.~Teasley.
\newblock A time series interaction analysis method for building predictive
  models of learners using log data.
\newblock In {\em Proceedings of the Fifth International Conference on Learning
  Analytics And Knowledge}, LAK '15, pages 126--135, New York, NY, USA, 2015.
  ACM.

\bibitem{brown15w}
R.~Brown, C.~Lynch, M.~Eagle, J.~Albert, T.~Barnes, R.~S. Baker, Y.~Bergner,
  and D.~S. McNamara.
\newblock Good communities and bad communities: Does membership affect
  performance?
\newblock In {\em Proceedings of the 8th International Conference on
  Educational Data Mining, {EDM} 2015, Madrid, Spain, June 26-29, 2015}, pages
  612--613, 2015.

\bibitem{brown15}
R.~Brown, C.~Lynch, Y.~Wang, M.~Eagle, J.~Albert, T.~Barnes, R.~S. Baker,
  Y.~Bergner, and D.~S. McNamara.
\newblock Communities of performance \& communities of preference.
\newblock In {\em EDM (Workshops)}, 2015.

\bibitem{chen17}
Y.~Chen and M.~Zhang.
\newblock Mooc student dropout: Pattern and prevention.
\newblock In {\em Proceedings of the ACM Turing 50th Celebration Conference -
  China}, ACM TUR-C '17, pages 4:1--4:6, New York, NY, USA, 2017. ACM.

\bibitem{dash1997feature}
M.~Dash and H.~Liu.
\newblock Feature selection for classification.
\newblock {\em Intelligent data analysis}, 1(3):131--156, 1997.

\bibitem{eckles12}
J.~E. Eckles and E.~G. Stradley.
\newblock A social network analysis of student retention using archival data.
\newblock {\em Social Psychology of Education}, 15(2):165--180, 2012.

\bibitem{fei15}
M.~Fei and D.-Y. Yeung.
\newblock Temporal models for predicting student dropout in massive open online
  courses.
\newblock In {\em Data Mining Workshop (ICDMW), 2015 IEEE International
  Conference on}, pages 256--263. IEEE, 2015.

\bibitem{freeman}
L.~C. Freeman.
\newblock Centrality in social networks conceptual clarification.
\newblock {\em Social networks}, 1(3):215--239, 1978.

\bibitem{gitinabard17w}
N.~Gitinabard, L.~Xue, C.~Lynch, S.~Heckman, and T.~Barnes.
\newblock A social network analysis on blended courses.
\newblock In {\em Proceedings of the 10th International Conference on
  Educational Data Mining, {EDM} 2017(Workshops), Wuhan, China, June 25-28,
  2017}, 2017.

\bibitem{gutl14}
C.~G{\"u}tl, R.~H. Rizzardini, V.~Chang, and M.~Morales.
\newblock Attrition in mooc: Lessons learned from drop-out students.
\newblock In {\em International Workshop on Learning Technology for Education
  in Cloud}, pages 37--48. Springer, 2014.

\bibitem{halawa14}
S.~Halawa, D.~Greene, and J.~Mitchell.
\newblock Dropout prediction in moocs using learner activity features.
\newblock {\em Experiences and best practices in and around MOOCs}, 7:3--12,
  2014.

\bibitem{jiang14a}
S.~Jiang, S.~M. Fitzhugh, and M.~Warschauer.
\newblock Social positioning and performance in moocs.
\newblock In {\em Workshop on Graph-Based Educational Data Mining}, volume~14,
  2014.

\bibitem{jiang14b}
S.~Jiang, A.~Williams, K.~Schenke, M.~Warschauer, and D.~O'dowd.
\newblock Predicting mooc performance with week 1 behavior.
\newblock In {\em Educational Data Mining 2014}, 2014.

\bibitem{joksimovic15}
S.~Joksimovi{\'c}, D.~Ga{\v{s}}evi{\'c}, V.~Kovanovi{\'c}, B.~E. Riecke, and
  M.~Hatala.
\newblock Social presence in online discussions as a process predictor of
  academic performance.
\newblock {\em Journal of Computer Assisted Learning}, 31(6):638--654, 2015.

\bibitem{jordan14}
K.~Jordan.
\newblock Initial trends in enrolment and completion of massive open online
  courses.
\newblock {\em The International Review of Research in Open and Distributed
  Learning}, 15(1), 2014.

\bibitem{kartsonaki2016survival}
C.~Kartsonaki.
\newblock Survival analysis.
\newblock {\em Diagnostic Histopathology}, 22(7):263--270, 2016.

\bibitem{kleinberg}
J.~M. Kleinberg.
\newblock Authoritative sources in a hyperlinked environment.
\newblock {\em J. ACM}, 46(5):604--632, Sept. 1999.

\bibitem{kloft14}
M.~Kloft, F.~Stiehler, Z.~Zheng, and N.~Pinkwart.
\newblock Predicting mooc dropout over weeks using machine learning methods.
\newblock In {\em Proceedings of the EMNLP 2014 Workshop on Analysis of Large
  Scale Social Interaction in MOOCs}, pages 60--65, 2014.

\bibitem{miller11}
R.~G. Miller~Jr.
\newblock {\em Survival analysis}, volume~66.
\newblock John Wiley \& Sons, 2011.

\bibitem{pursel16}
B.~Pursel, L.~Zhang, K.~Jablokow, G.~Choi, and D.~Velegol.
\newblock Understanding mooc students: Motivations and behaviours indicative of
  mooc completion.
\newblock {\em J. Comp. Assist. Learn.}, 32(3):202--217, June 2016.

\bibitem{rose14}
C.~P. Ros{\'e}, R.~Carlson, D.~Yang, M.~Wen, L.~Resnick, P.~Goldman, and
  J.~Sherer.
\newblock Social factors that contribute to attrition in moocs.
\newblock In {\em Proceedings of the first ACM conference on Learning@ scale
  conference}, pages 197--198. ACM, 2014.

\bibitem{sinha14}
T.~Sinha, N.~Li, P.~Jermann, and P.~Dillenbourg.
\newblock Capturing" attrition intensifying" structural traits from didactic
  interaction sequences of mooc learners.
\newblock {\em arXiv preprint arXiv:1409.5887}, 2014.

\bibitem{taylor14}
C.~Taylor et~al.
\newblock {\em Stopout prediction in massive open online courses}.
\newblock PhD thesis, Massachusetts Institute of Technology, 2014.

\bibitem{vihavainen13}
A.~Vihavainen, M.~Luukkainen, and J.~Kurhila.
\newblock Using students' programming behavior to predict success in an
  introductory mathematics course.
\newblock In {\em Educational Data Mining 2013}, 2013.

\bibitem{wang14}
Y.~Wang.
\newblock Mooc leaner motivation and learning pattern discovery.
\newblock In {\em EDM}, pages 452--454, 2014.

\bibitem{yang16}
D.~Yang, R.~Kraut, and C.~P. Ros{\'e}.
\newblock Exploring the effect of student confusion in massive open online
  courses.
\newblock {\em Journal of Educational Data Mining}, 8(1), 2016.

\bibitem{yang13}
D.~Yang, T.~Sinha, D.~Adamson, and C.~P. Ros{\'e}.
\newblock Turn on, tune in, drop out: Anticipating student dropouts in massive
  open online courses.
\newblock In {\em Proceedings of the 2013 NIPS Data-driven education workshop},
  volume~11, page~14, 2013.

\bibitem{zhu16}
M.~Zhu, Y.~Bergner, Y.~Zhang, R.~Baker, Y.~Wang, and L.~Paquette.
\newblock Longitudinal engagement, performance, and social connectivity: a mooc
  case study using exponential random graph models.
\newblock In {\em Proceedings of the Sixth International Conference on Learning
  Analytics \& Knowledge}, pages 223--230. ACM, 2016.

\end{thebibliography}
\end{document}